\documentclass[fleqn,usenatbib]{mnras}

\usepackage{newtxtext,newtxmath}

\usepackage[T1]{fontenc}

\DeclareRobustCommand{\VAN}[3]{#2}
\let\VANthebibliography\thebibliography
\def\thebibliography{\DeclareRobustCommand{\VAN}[3]{##3}\VANthebibliography}

\usepackage{graphicx}	
\usepackage{amsmath}	

\usepackage{ulem}
\usepackage{color}
\usepackage{xcolor}

\usepackage{tikz,xcolor,hyperref}

\definecolor{lime}{HTML}{A6CE39}
\DeclareRobustCommand{\orcidicon}{%
	\begin{tikzpicture}
	\draw[lime, fill=lime] (0,0) 
	circle [radius=0.16] 
	node[white] {{\fontfamily{qag}\selectfont \tiny ID}};
	\draw[white, fill=white] (-0.0625,0.095) 
	circle [radius=0.005];
	\end{tikzpicture}
	\hspace{-2mm}
}

\foreach \x in {A, ..., Z}{%
	\expandafter\xdef\csname orcid\x\endcsname{\noexpand\href{https://orcid.org/\csname orcidauthor\x\endcsname}{\noexpand\orcidicon}}
}

\title[]{Galactic cosmic ray transport in the absence of resonant scattering}

\author[O. Pezzi \& P. Blasi]{
O. Pezzi\orcidA{}$^{1}$\thanks{E-mail: oreste.pezzi@istp.cnr.it} and
P. Blasi\orcidB{}$^{2,3}$
\\
$^{1}$Istituto per la Scienza e Tecnologia dei Plasmi, Consiglio Nazionale delle Ricerche, Via Amendola 122/D, I-70126 Bari, Italy\\
$^{2}$Gran Sasso Science Institute, Viale F. Crispi 7, I-67100 L'Aquila, Italy\\
$^{3}$INFN/Laboratori Nazionali del Gran Sasso, Via G. Acitelli 22, I-67100 Assergi (AQ), Italy}

\date{Accepted XXX. Received YYY; in original form ZZZ}

\pubyear{2023}

\begin{document}
\label{firstpage}
\pagerange{\pageref{firstpage}--\pageref{lastpage}}
\maketitle

\begin{abstract}
Galactic cosmic ray transport relies on the existence of turbulence on scales comparable with the gyration radius of the particles and with wavenumber vector oriented along the local magnetic field. In the standard picture, in which turbulence is injected at large scales and cascades down to smaller scales, it is all but guaranteed that the turbulent fluctuations at the scales relevant for resonant scattering may be present,
either because of anisotropic cascading or because of the onset of damping processes. This raises questions on the nature of cosmic-ray scattering, especially at energies $\gtrsim 1$ TeV, where self-generation is hardly relevant. Here, by means of numerical simulations of charged test-particles in a prescribed magnetic field, we perform a {\it gedankenexperiment} aimed at investigating particle diffusion in a situation in which turbulence is mainly present at large scales, and discuss possible implications of this setup for cosmic-ray transport phenomenology.
\end{abstract}




\section{Introduction}
\label{sect:intro}

The simplest description of cosmic ray (CR) transport in the Galaxy relies on the existence of possibly-broadened resonances between particle gyration and $1/k$, $k$ being the wavenumber of turbulent fluctuations. At first glance, the approximately power-law-like shape of the CR energy spectrum over many decades well reconciles with the power-law turbulent spectrum in $k$-space, as turbulence is injected on large scales and eventually cascades to smaller spatial scales. 

This simple picture has been recently challenged based on both observational and theoretical considerations. The AMS-02 measurements identified a spectral hardening at rigidity $\sim 200-300$ GV in the spectrum of CR protons \citep{Choutko_2016}, helium \citep{Aguilar_He}, and heavier nuclei \citep{Aguilar_2017}. This feature was also observed in the secondary to primary ratios, such as B/C and B/O \citep{Aguilar_BC}, thus strongly suggesting that the slope change is due to a variation of the diffusive properties of CRs in the Galaxy at the same rigidity \citep{Genolini_2017, evoli2020ams}. Such a finding indicates that either the scattering properties or the nature of turbulence change at $\sim 200$ GV. 
The spectral hardening has been associated with either a non-separable spatial dependence of the diffusion coefficient in the Galaxy \citep{Tomassetti2012} or the transition from self-generated to pre-existing turbulence \citep{Blasi2012}. As discussed by \cite{Kulsrud_1969,Skilling_1975,Holmes_1975}, CRs with rigidity below a few hundred GV can create their own scattering centers, while at higher energies some pre-existing Galactic turbulence, injected on large scales and cascading down to smaller spatial scales, is needed \citep{Blasi2012,2018AdSpR..62.2731A}, though the possibility that scattering may be self-generated at all energies has been recently put forward \citep{Dogiel2022}.

In the simple scenario illustrated above, the match between the two regimes typically occurs in the few hundred GV rigidity range, if the cascade is assumed to be isotropic in $k$ space \citep{Zhou_1990}. This latter assumption is however at odds with the spectral anisotropy of Alfv\'enic turbulence \citep{shebalin1983anisotropy, sridhar1994toward, goldreich1995toward, matthaeus1996anisotropic, matthaeus2012local, oughton2020critical,scheko2022MHD}. Indeed, most of the turbulent power is efficiently transferred to smaller scales in the directions perpendicular to the magnetic field rather than in the parallel direction, and the pitch-angle diffusion coefficient decreases as a result of this effect \citep{chandran2000scattering}. 

Magnetosonic fluctuations have also been considered to overcome the anisotropy issue \citep{cho2002compressible, cho2002simulations}. In this context, the spatial diffusion coefficient is usually calculated from the pitch-angle diffusion coefficient $D_{\mu\mu}$ in the quasi-linear/weakly-nonlinear framework \citep{volk1973nonlinear, chandran2000scattering, yan2002scattering, yan2008cosmicray,fornieri2021theory}. However, magnetosonic modes have their own problematic aspects. Slow magnetosonic fluctuations cascade anisotropically in $k$-space, similar to Alfv\'enic ones. On the other hand, the cascade of fast magnetosonic modes is approximately isotropic, but their spectrum is affected by various damping mechanisms, whose relevance critically depends on the medium properties, such as the plasma $\beta$ (kinetic to magnetic pressure ratio). Indeed, in older literature, fast modes were never considered in the context of CR diffusion since they were expected to be heavily damped \citep{barnes1966collisionless}. Moreover, the inferred diffusion coefficient due to scattering off magnetosonic modes strongly depends on the plasma $\beta$, thus making a reliable computation hard to achieve (see, e.g., \cite{fornieri2021theory} for a recent calculation), and has odd energy dependence --often implying superluminal particle motion in the parallel direction-- for a wide choice of parameters. Recently, \citet{Kempski_2022} argued that fast magnetosonic modes may develop shocklets while cascading, thus leading to unusual CR scattering properties. All these calculations proceed along the same line: one computes the parallel diffusion coefficient and argues that this reflects the general scattering properties. We argue that this second implication is, in general, incorrect in that the global diffusion properties are sensitive to the random walk of magnetic field lines (FLRW), which itself forces the direction of motion of CRs to change. Again, the issue of high-energy ($\gtrsim$ TeV) CR scattering appears all but settled.

Aside from these theoretical approaches, the spatial diffusion coefficient has also been computed numerically by propagating charged test-particles in synthetic models of turbulence either generated on grid \citep{demarco2007numerical, dundovic2020novel, reichherzer2022anisotropic, kuhlen2022field} or built as a superposition of plane waves \citep{casse2001transport, pucci2016energetic, mertsch2020test}. Most notably, there have been attempts at obtaining clues on CR transport by directly propagating particles in turbulent fields directly obtained through magnetohydrodynamic (MHD) simulations \citep{beresnyak2011numerical, cohet2016cosmic, pezzi2022relativistic}. This latter approach may appear as the most appropriate for assessing the role of different MHD modes. However, the dynamic range achieved in MHD simulations is typically rather small, covering only about 3 decades in $k$-space, of which only one and a half below the energy containing scale. This limitation sets a correspondingly small energy range for which the CR transport is properly described. Such a dynamic range is too small to explore to its full extent the anisotropic cascade of Alfv\'enic and slow magnetosonic modes (see, e.g., \citet{cohet2016cosmic, pezzi2022relativistic}). 

The confusing situation illustrated above becomes even more puzzling if one recalls that Galactic CRs spend most of their confinement time in the Galactic halo. In this region, little fresh injection of turbulence on large scales is expected, although galaxy mergers \cite[]{Setton2023} and winds/outflows may provide some contribution to halo turbulence. The halo turbulence might also originate in the disc region and eventually be advected into the halo while cascading from the injection scale $L$ to smaller scales, on times of the order of $L/v_A\sim$ few million years, being  $v_A=B/\sqrt{4\pi\rho}$ the Alfv\'en speed computed with the typical magnetic field $B$ and plasma density $\rho$. Note that such a time scale is rather close to the confinement time of TeV CRs and it is not guaranteed that the cascade may develop completely.

All these concerns and, particularly, the anisotropy of Alfv\'enic turbulence and the possible damping mechanisms affecting both Alfv\'enic and fast magnetosonic fluctuations, have motivated the thought experiment underlying the present work. Our goal is to investigate the CR transport in the Galaxy in a scenario in which the turbulent cascade does not develop or develops partially, so that most of the power remains in a narrow neighborhood of the coherence length $l_c$,  while little power is available at scales $\ll l_c$. One can picture this magnetic field configuration as a coherent field whose orientation changes in cells of size $\sim l_c$. In the community involved in the investigation of the Galactic magnetic field through Faraday rotation measures, this would correspond to the {\it ordered field} \citep{Ferriere2020}, while the denomination {\it turbulent field} is limited to smaller-scale fluctuations.

In this scenario, particle transport results from the combined effect of the ``standard'' resonant scattering and FLRW \citep{jokipii1969stochastic}, the latter determining particle transport in the absence of turbulent cascade. FLRW has been extensively investigated to characterize magnetic field topology and trapping \citep{matthaeus1995spatial, chuychai2007trapping, sonsrette2016magnetic} and it is known to influence particle diffusion across field lines \citep{chuvilgin1993anomalous, giacalone1999transport, casse2001transport, webb2006compound} (see also \citet{shalchi2020perpendicular} and references therein and \citet{kuhlen2022field}). The effect of the FLRW on the electron heat conduction has also received notable attention \citep{chandran1999thermal, lazarian2006enhancement}. Here, we discuss the implications of FLRW on the CR transport in the Galaxy, in the absence of prominent power on the resonant scales, mimicking either the anisotropic development of the Alfvenic/slow magnetosonic cascade or the damping of fast modes. This setup, when treated in the standard way, namely excluding the effect of FLRW, would lead to exceedingly large parallel pathlength (see, e.g., \citet{fornieri2021theory}).

The problem is investigated by constructing a three-dimensional (3D) magnetic field on grid and propagating charged test particles in such a field, without homogeneous, background component $B_0$. By changing the power in the form of large-scale coherent field with respect to the one in the form of turbulence, we modulate the role of FLRW. In this clearly idealized situation in which there is no power in the form of resonant modes, the confinement time of CRs on Galactic scales remains of easy physical interpretation and does not show pathological aspects such as superluminal particle motion.
In order to make more realistic predictions on the energy dependence of the confinement time, one should investigate numerous additional effects, such as multiple scales of turbulent injection, transition from self-generated turbulence and the likely presence of a background magnetic field, that are beyond the scope of this article.

\section{Numerical method and setups}
\label{sect:num}

We follow the trajectories of $N_p=8192$ relativistic test protons in a non-relativistic magnetostatic synthetic turbulence ($\partial {\bf B}/\partial t=0$, ${\bf E}=0$) in the absence of background field, $\langle {\bf B}\rangle=0$ being $\langle \dots \rangle$ here the average on the computational box. The motion equations are:
\begin{equation}
 \label{eq:lorentz}
 \begin{split}
   \frac{d {\bf r}}{d t} &= {\bf v} \\
     \frac{d {\bf p}}{d t} &= e\ \frac{{\bf v}}{c}\times {\bf \delta B}({\bf r})\,,
 \end{split}
\end{equation}
being $e$ the proton charge and $c$ the speed of light. Particle position, speed, momentum, and Lorentz factor are respectively indicated by ${\bf r}$, ${\bf v}$, ${\bf p}=m_p\gamma{\bf v}$, and $\gamma=\sqrt{1+{\bf p}^2/m_p^2c^2}$, being $m_p$ the proton mass. The purely-fluctuating magnetic field ${\bf \delta B}({\bf r})$, with r.m.s. amplitude $\delta B_{\rm rms} = \sqrt{\langle{\bf \delta B}^2\rangle}$, is split as ${\bf \delta B}({\bf r}) = {\bf \delta B}_{\rm res}({\bf r}) + {\bf \delta B}_{\rm coh}({\bf r})$, being $\langle{\bf \delta B}_{\rm res}\rangle = \langle{\bf \delta B}_{\rm coh}\rangle = 0$, while their r.m.s. amplitudes satisfy $\delta B_{\rm rms}^2= \delta B_{\rm res}^2 + \delta B_{\rm coh}^2=1 {\rm \mu G}$. ${\bf \delta B}_{\rm res}({\bf r})$ is the pure turbulent component modeled here with a Kolmogorov-like spectrum and responsible for resonant scattering of lower energy particles. ${\bf \delta B}_{\rm coh}({\bf r})$ mimics the large-scale complexity of the magnetic field responsible for FLRW.  

The spectrum function $S(k)=S^{\rm res}(k)+S^{\rm coh}(k)$, associated with ${\bf \delta B}({\bf r})$, is composed of:
\begin{equation}
  \begin{split}
     S^{\rm res}(k) &=  \delta B_{\rm res}^2 \frac{C(q=4,s)}{\pi k^2} l_{\rm iso} \frac{\left(k l_{\rm iso}\right)^4}{\left(1 + k^2 l_{\rm iso}^2 \right)^{s/2+2}} ,\\ 
    S^{\rm coh}(k) &= \delta B_{\rm coh}^2 \frac{\left(k/\Delta k^*\right)^2}{(2\pi)^{3/2}{\Delta k^*}^3 D(k^*/\Delta k^*)} \exp\left[{-\frac{1}{2}\left(\frac{k-k^*}{\Delta k^*}\right)^2}\right].
  \end{split}
  \label{eq:S}
\end{equation}
In the expression for $S^{\rm res}(k)$, the bend-over scale $l_{\rm iso}$ is the scale at which the inertial range of turbulence, whose slope is controlled by the parameter $s=5/3$, begins. Here, we set $L_{\rm box}/l_{\rm iso} = 8$ to include several correlation lengths in the box, with $L_{\rm box}=512 {\rm pc}$. The normalization constant $C(q,s)=\Gamma\left(\frac{s+q}{2}\right)/\left(2 \Gamma\left(\frac{s-1}{2}\right)\Gamma\left(\frac{q+1}{2}\right)\right)$, being $\Gamma$ the Euler gamma function, ensures that the energy density is normalized to $\delta B_{\rm res}^2$, i.e., $\delta B_{\rm res}^2= 4 \pi \int_0^\infty dk k^2 S^{\rm res}(k)$, with $q=4$ to guarantee the homogeneity of turbulent fluctuations \citep{batchelor1982,matthaeus2007spectral,sonsrettee2015magnetic,dundovic2020novel}.

\begin{figure}
\centering
	\includegraphics[width=0.92\columnwidth]{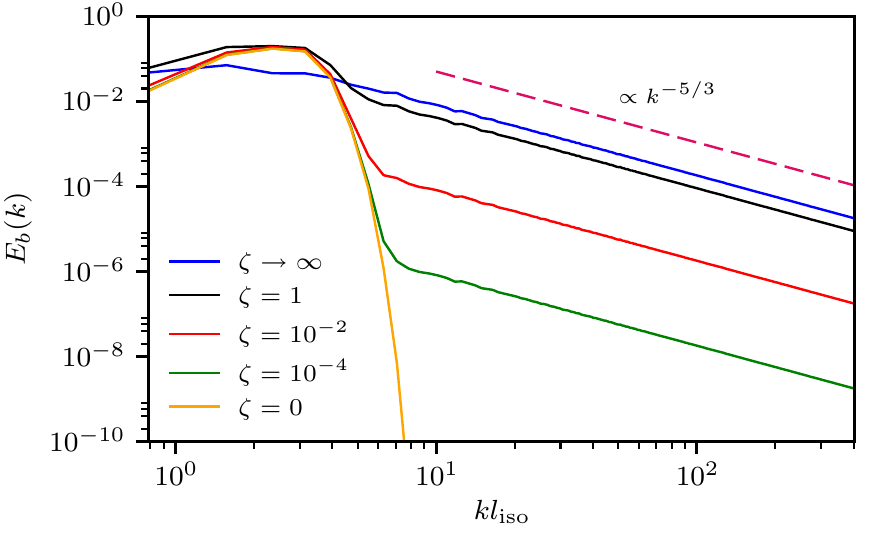}
    \caption{Omni-directional spectrum of the magnetic field energy, $E_b(k)$, for several values of $\zeta$: $\zeta\rightarrow\infty$ (blue), $\zeta=1$ (black), $\zeta=10^{-2}$ (red), $\zeta=10^{-4}$ (green), and $\zeta=0$ (orange). The cerise-red dashed line displays the Kolmogorov $k^{-5/3}$ slope for reference. }
    \label{fig:Spectra}
\end{figure}

$S^{\rm coh}(k)$ represents a Gaussian bump in $k$-space centered at $k=k^*=2/l_{\rm iso}$ whose width is $\Delta k^*=\pi/(4 l_{\rm iso})$ and such that it preserves homogeneity. By normalizing the energy density to $\delta B_{\rm coh}^2$, one finds $D(x)=3+8\sqrt{2/\pi}x+6x^2+4\sqrt{2/\pi}x^3+x^4$, being $x=k^*/\Delta k^*$.

With the spectral function $S(k)$ in Eqs. (\ref{eq:S}), the correlation length of the magnetic field fluctuations can be estimated from Eq. (45) of \citet{sonsrettee2015magnetic} by assuming the random ballistic decorrelation closure for the magnetic field lines:
\begin{equation} 
  \frac{l_c}{l_{\rm iso}} = \frac{2 \pi}{1 + \zeta} \left[\zeta \frac{2 C(q=4,s)}{s(s+2)} + \frac{1}{2\sqrt{2\pi}} \cdot \frac{D'(k^*/\Delta k^*)}{D(k^*/\Delta k^*) \left(l_{\rm iso}\Delta k^*\right)} \right]
  \label{eq:lc}
\end{equation}
where $\zeta=\delta B_{\rm res}^2/\delta B_{\rm coh}^2$ and $D'(x)=2+3\sqrt{\pi/2}x+3 x^2 + \sqrt{\pi/2} x^3$, being again $x=k^*/\Delta k^*$. If $\zeta\rightarrow\infty$, Eq. (\ref{eq:lc}) reduces to Eq. (21) of \citet{dundovic2020novel}. With our choice of the parameters $l_{\rm iso}$, $k^*$, and $\Delta k^*$, we get $l_c\simeq 0.5 l_{\rm iso}=32 {\rm pc}$ for different values of $\zeta$. Such a correlation length, in line with values inferred from observations \cite[]{Beck2015}, corresponds to the gyroradius of a particle with energy $E(r_g=l_c)\simeq 30 {\rm PeV}$. The above expression is in agreement with the correlation length directly computed from the magnetic field at different $\zeta$, being the discrepancy always $\lesssim5\%$. One should keep in mind that the numerical value of $l_c$ in our simulations should be interpreted as a sort of spatially-averaged value of the correlation length over the diffusion volume, a quantity that we have poor access on. As an instance, spatial variations in the coherence scale are observed in the solar wind \citep{ruiz2014characterization}.

The different role of ${\bf \delta B}_{\rm res}({\bf r})$ and ${\bf \delta B}_{\rm coh}({\bf r})$ can be appreciated from Figure \ref{fig:Spectra} that displays the magnetic field omni-directional energy spectrum $E_b(k)$ for different values of $\zeta$. As $\zeta$ decreases, less power is available for resonant scattering and goes into the coherent field component. The case $\zeta\rightarrow\infty$ corresponds to the one explored by \citet{dundovic2020novel} as the power is entirely associated to a Kolmogorov-like turbulent component. In turn, the case $\zeta=0$ implies that all power is contained in a coherent field that changes direction on a scale $\sim 1/k^*\sim l_c$. 

Magnetic fluctuations are generated on a tri-periodic cubic grid of size $L_{\rm box}$ discretized with $N=1024$ grid points in each direction, with double-precision floating point. The spectral tensor of the magnetic fluctuations is assigned in the wavenumber space that is limited, in each direction, by $k_{\rm min}=k_0=2\pi/L_{\rm box}=\pi/4 l_{\rm iso}$ and the Nyquist wave number $k_{\rm max}= k_0 (N/2-1)$, following the procedure proposed by  \citet{sonsrettee2015magnetic} which fulfills $\nabla \cdot {\bf \delta B}({\bf r})=0$.  
Particle trajectories are obtained by integrating Eqs. (\ref{eq:lorentz}) with the symplectic Boris method, where tri-linear interpolation method is adopted to evaluate the magnetic field at the particle position. 
For one set of parameters, the integration is repeated for an ensemble of $N_p$ particles in a single realization of the magnetic field \citep{dundovic2020novel}. 

\begin{figure}
\centering
	\includegraphics[width=0.92\columnwidth]{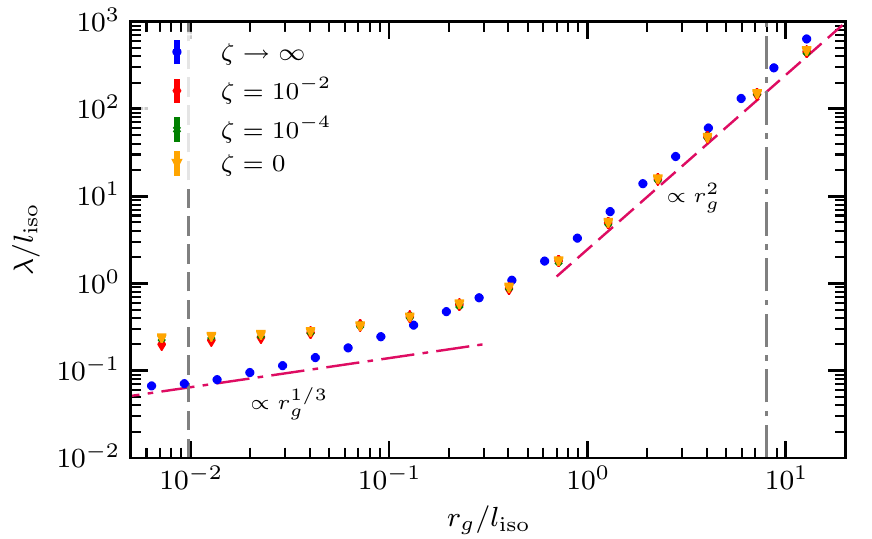}
    \caption{Isotropic mean free path $\lambda=3D_{\rm iso}/c$ as a function of the particle gyroradius $r_g$, for $\zeta\rightarrow\infty$ (blue dots), $\zeta=10^{-2}$ (red rhombuses), $\zeta=10^{-4}$ (green crosses), and $\zeta=0$ (orange triangles). Cerise-red lines respectively display the $r_g^{1/3}$ dependence (dot-dashed) and the $r_g^{2}$ dependence (dashed). Gray vertical lines indicate the gyroradius at which resonance is lost for numerical reasons (dashed) and corresponding to the box size $r_g=L_{\rm box}$ (dot-dashed), respectively.  }
    \label{fig:Dsat}
\end{figure}

\section{Simulation results}
\label{sect:sim}
An extensive numerical campaign has been conducted to tackle the following two scientific questions.

\textit{How do particles diffuse in the absence of turbulence at resonant scales?} We carried out a series of simulations in which we change the value of $\zeta$. For each value of $\zeta$ we performed several runs at different particle gyroradius in the range $r_g/l_{\rm iso}\in[10^{-2},10^1]$, i.e., $E_{\rm PeV}\in[0.6, 6 \times 10^{2}]$. Particles are injected homogeneously throughout the box with speed directions sampled uniformly on the 3D sphere. Given the grid size of $1024^3$, the chosen gyroradius range guarantees to describe particle scattering in about one energy decade in the inertial range of turbulence and to recover the high-energy transition occurring at scales about $r_g\sim l_c$ \citep{subedi2017charged,dundovic2020novel}. For each run, we calculated the spatial diffusion coefficient along each direction, e.g., $D_{\rm xx}$, from the time-dependent diffusion coefficients $D^{\rm run}_{\rm xx}(t) = \langle (\Delta x(t))^2 \rangle/2t$, being $\Delta x(t)$ the particle displacement during a time interval $t$ and $\langle ... \rangle$ indicating hereafter the average on the particle ensemble. $D_{\rm xx}$ is the long-time limit of $D^{\rm run}_{\rm xx}(t)$ when it has reached the diffusive plateau. Analogous definitions hold for the $y$ and $z$ directions. For a vanishing regular field, particle diffusion is spatially isotropic and the particle mean free path is $\lambda = 3 D_{\rm iso}/c$, being $D_{\rm iso}=(D_{\rm xx}+D_{\rm  yy}+D_{\rm  zz})/3$ the isotropic diffusion coefficient.

Figure \ref{fig:Dsat} shows the mean free path $\lambda$ as a function of the particle gyroradius $r_g$ for $\zeta\rightarrow\infty$ (blue dots), $\zeta=10^{-2}$ (red rhombuses), $\zeta=10^{-4}$ (green crosses), and $\zeta=0$ (orange triangles). We disregarded the case $\zeta=1$ shown in Figure \ref{fig:Spectra} since the magnetic energy spectrum is quite similar to the one achieved in the $\zeta\rightarrow\infty$ case. At high energies, i.e., $r_g/l_{\rm iso} \gtrsim 1$, the transition to the scaling $\lambda_{\rm iso} \sim r_g^2$ (cerise-red dashed line in Fig. \ref{fig:Dsat}) is recovered for all values of $\zeta$.

In the case $\zeta\rightarrow\infty$, the power spectrum changes continuously in $k$, namely there is no jump at $k>1/l_c$. Hence, the corresponding pathlength for particle scattering also changes in a continuous manner with energy and its energy dependence agrees, for gyroradius $r_g<l_c$, with the prediction $\sim r_g^{1/3}$ (cerise-red dot-dashed line in Fig. \ref{fig:Dsat}). This is not the case when $\zeta\ll 1$. For particles with energy below $E(l_c)$, the pathlength due to resonant scattering would become $\gg l_c$, hence transport is dominated by FLRW, i.e., the pathlength becomes independent of energy and $\sim l_c$ (recall that $l_c/l_{\rm iso}\sim 0.5$). This is clear from inspecting the low-energy behaviour of the curves in Figure \ref{fig:Dsat} with $\zeta \ll 1$. We expect that for any value of $\zeta$ there should be a sufficiently small energy for which the pathlength due to resonant scattering becomes again smaller than $l_c$: at such energies one should recover the energy dependent pathlength expected in quasi-linear theory and previously found by \citep{subedi2017charged, dundovic2020novel}. Given the dynamic range of our simulations, it is difficult to test this expectation here. 

This situation qualitatively resembles what is expected in the presence of self-generation, which produces fluctuations only at large wavenumbers. Indeed, if we assume that the diffusion coefficient at $R<TV$ as inferred from secondary to primary ratios is $D(E)\sim 3\times 10^{28}E_{\rm GeV}^{1/2}~\rm cm^2/s$, then the corresponding pathlength becomes comparable with $l_c\simeq 32 \rm pc$ for energies of order $\sim \rm TeV$. Clearly our approach is inadequate to include self-generation and this numerical coincidence should only be taken as a hint that the CR confinement due to FLRW occurs on scales that are very likely to be relevant and should not be neglected.

We stress once more that the transport of high-energy particles on which these considerations are based has been obtained by requiring only that the large-scale magnetic field is randomly oriented on scales $\sim l_c$ and that there is no homogeneous background magnetic field. Hence, our numerical experiments provide a proof of principle that particle diffusion would be guaranteed even in pathological cases in which the Alfv\'enic turbulence becomes strongly anisotropic or in the presence of severe damping of fast magnetosonic modes. All these considerations turn out to be foundational for tackling the second following question.  

\textit{What is the confinement time the Galaxy in the absence of turbulent power on resonant scales?} In the energy range accessible to our simulations, the diffusion coefficient in the low-energy regime is roughly the same for small values of $\zeta$. Hence, we will focus here only on the limit cases $\zeta \rightarrow \infty$ and $\zeta=0$. For each value of $\zeta$, we performed different runs at different particle gyroradius in the range $r_g/l_{\rm iso}\in[10^{-2},10^2]$, i.e., particle energy $E_{\rm PeV}\in[0.6, 6 \times 10^{3}]$ in physical units. We extended this range for about a decade at high energies to recover the transition to ballistic transport. 

\begin{figure}
\centering
  \includegraphics[width=0.92\columnwidth]{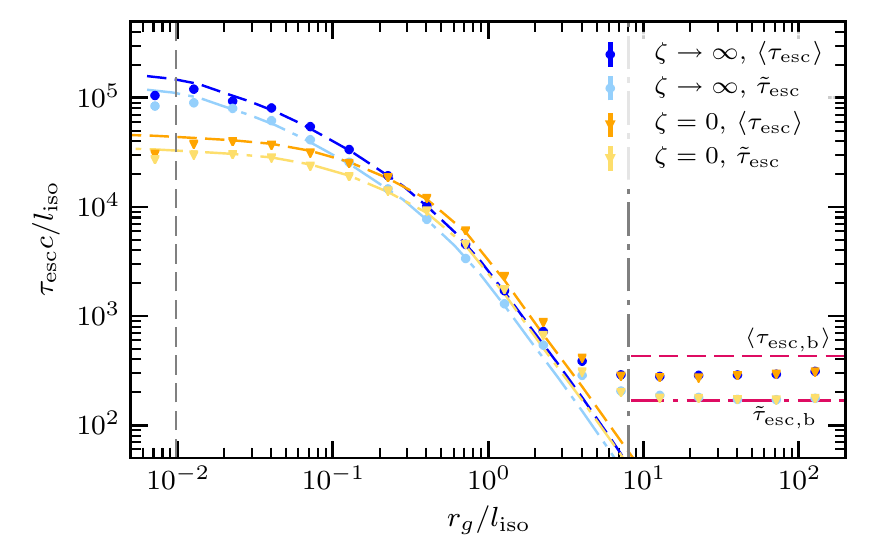}
    \caption{Escape times as a function of the particle gyroradius $r_g$. Blue (light blue) points indicate respectively the average (median) escape times, computed for the $\zeta\rightarrow\infty$ case. Orange (light orange) triangles refer to the average (median) escape times in the $\zeta=0$ case. Blue and orange dashed lines indicate the diffusive escape time $\tau_{\rm D}=L_{\rm esc}^2/2 D_{\rm iso}$ for the two cases $\zeta\rightarrow\infty$ and $\zeta=0$.
    Light blue and orange dot-dashed lines show the prediction $0.75\tau_{\rm D}$ for the two different setups. The horizontal cerise-red lines correspond to the average ballistic escape time, $\langle \tau_{\rm esc, b}\rangle$ (dashed) and the median ballistic escape time $\tilde{\tau}_{\rm esc,b}$ (dot dashed). Gray vertical lines indicate the gyroradius at which resonance is lost for numerical reasons, i.e. the gyroradius that comprises at least five grid points (dashed), and corresponding to the box size $r_g=L_{\rm box}$ (dot-dashed), respectively.}
    \label{fig:Tescape}
\end{figure}

For mimicking the particle escape from the Galactic halo, we identify the $(x,y)$ plane at $z=L_{\rm box}/2$ as the Galactic disc, while the halo fills the $z$ direction out to an escape boundary at $L_{\rm esc}= L_{\rm box}/2 + 10 L_{\rm box}$. Particles are injected isotropically in such an infinitesimally thin disc. A particle is considered as escaped when its displacement along the $z$ direction reaches the threshold height $L_{\rm esc}$. In order to simulate this situation, we exploit periodic boundary conditions on the individual box, while keeping track on the displacement in the $z$ direction. At the end of each run, all particles escape. Due to the randomness of the magnetic field perturbations, the probability distribution function (PDF) of the escape times is wide (not shown here), thus indicating that particles escape with different times $\tau_{{\rm esc}}$. The average, $\langle \tau_{{\rm esc}} \rangle$, and median, $\tilde{\tau}_{{\rm esc}}$, escape times are evaluated at the end of each run. 

Figure \ref{fig:Tescape} shows the energy dependence of escape times. Blue and orange colors refer to the cases $\zeta\rightarrow \infty$ and $\zeta=0$, respectively. Blue and orange markers show the average escape times $\langle \tau_{\rm esc}\rangle$, while light blue and orange markers display the median escape time $\tilde{\tau}_{\rm esc}$. Blue and orange dashed lines indicate the average diffusive escape time $\tau_{D} = L_{\rm esc}^2/2 D_{\rm iso}$ (see, e.g., Eqs. (11) and (20-22) of \citet{lipari2014lifetime}), while light blue and orange dot-dashed lines display the prediction for the median diffusive escape time $\simeq 0.75 \tau_{D}$, calculated from the distribution of Eqs. (20-21) in \citet{lipari2014lifetime}. In both cases, the isotropic diffusion coefficient $D_{\rm iso}$ has been computed directly from simulations as in Figure \ref{fig:Dsat}.

Several interesting features emerge by inspecting Figure \ref{fig:Tescape}. At most energies, particle escape is well described by the diffusion approximation, as expected since $\lambda \ll L_{\rm esc}$ (Fig. \ref{fig:Dsat}). The different diffusion properties recovered at low energies for decreasing $\zeta$ values reflect into a distinct energy dependence of the escape times. In the $\zeta\rightarrow\infty$ case, the escape time shows a trend close to $r_g^{-1/3}$, while in the FLRW-dominated scenario, the escape time becomes constant as energy decreases. 

At high enough energies the diffusive approximation eventually breaks down,
and a transition to a ballistic escape is observed. For both $\zeta\rightarrow\infty$ and $\zeta=0$, the median and average escape times saturate at two different values, respectively close to $\tilde{\tau}_{\rm esc,b}=2 L_{\rm esc}/c$ (cerise-red dot-dashed) and $\langle \tau_{\rm esc, b}\rangle\simeq \frac{L_{\rm esc}}{c} \ln{\left(\frac{L_{\rm esc}}{l_c}\right)}$ (cerise-red dashed). Such a behavior can be interpreted as follows. 

As we inject $N_p$ particles isotropically, the number of particles moving with pitch angle $0\leq \mu\leq 1$, being $\mu$ evaluated with respect to the escape axis $z$, is $n(\mu)=N_p/2$. These particles escape in a time $\tau\simeq L_{\rm esc}/c\mu$. The number of particles with escape time between $\tau$ and $\tau+d\tau$ can be easily inferred as:
\begin{equation}
n(\tau)=\frac{N_p}{2} \frac{L_{\rm esc}}{c \tau^2} \, .
\end{equation}
The slope of the PDFs of $n(\tau)$ measured from numerical simulations is in agreement with the above $\simeq -2$ slope (not shown here). By requiring that the number of particles escaping with times $> \tau$ are half of the total number of particles with $\mu>0$, i.e., $N_p/4$, it is easy to compute the median ballistic escape time $\tilde{\tau}_{\rm esc,b}=2 L_{\rm esc}/c$. The average ballistic escape time is instead:
\begin{equation}
 \langle \tau_{\rm esc,b} \rangle= \frac{\int_{\tau_{\rm min}}^{\tau_{\rm max}} d\tau n(\tau) \tau}{ \int_{\tau_{\rm min}}^{\tau_{\rm max}} d\tau n(\tau)}\simeq \frac{L_{\rm esc}}{c} \ln{ \left(\frac{c \tau_{\rm max}}{L_{\rm esc}}\right)} \, ,
\end{equation}
where $\tau_{\rm min} =L_{\rm esc}/c$ and $\tau_{\rm max}\gg \tau_{\rm min}$ is related to the particles with very small pitch-angle, $\mu_{\min}$, that spend a rather long time in the halo. We estimated $\mu_{\min}$ by noticing that these particles perform a random walk deflecting by an angle $\Delta \theta \simeq \eta^{1/2} l_c/r_g$, where the number of scatterings is $\eta\simeq L_{\rm esc}/(\mu l_c)$. Since the particle gyration is only weakly perturbed, the particles should escape when $r_g \Delta \theta \sim L_{\rm esc}$, i.e., $\mu_{\rm min} \simeq l_c/L_{\rm esc}$, thus providing $\tau_{\rm max}\simeq L_{\rm esc}^2/(c l_c)$ and $\langle \tau_{\rm esc,b} \rangle \simeq \frac{L_{\rm esc}}{c} \ln{\left(\frac{ L_{\rm esc}}{l_c}\right)}$. This trend is shown in Figure \ref{fig:Tescape} with an horizontal dashed line and agrees well with the average escape time computed from simulations. 

\section{Discussion}
\label{sect:disc}

The recent measurement of features in the spectra of both primary and secondary cosmic rays has led to a revival in the interest for CR self-generation and for scattering in general (see  \cite{2018AdSpR..62.2731A} for a review). The nature of the scattering for particles with energy $\gtrsim$ TeV, for which self-generation is expected to be inefficient and pre-existing turbulence is needed to provide scattering centers, remains rather puzzling. Alfv\'enic and slow magnetosonic turbulence are known to cascade in an anisotropic way, leaving little power at large parallel wavenumbers where resonances would lead to particle scattering within simplified transport models that assume quasi-linear theory and random phases. Fast magnetosonic modes cascade more isotropically but the scattering they lead to is extremely sensitive to the plasma conditions, rather unlike the apparent homogeneity in the spectrum of CRs inferred from observation of diffuse gamma rays in the Milky Way. The fast modes might even be damped due to the formation of shocklets \citep{Kempski_2022}. This rather odd situation often reflects in unphysical diffusion coefficients, even implying super-luminal motion of CRs at high energies \cite{fornieri2021theory}. We argue that these conclusions are due to the misleading assumption that parallel resonant scattering is all what is relevant for CR confinement, thereby neglecting FLRW. The latter is dominated by the large-scale complexity of the magnetic field and is not affected by the way the cascade takes place. Hence, its effects is present even in the unlikely case that no power exists at the resonant scales.

Here, we investigated the problem of CR confinement in the Galaxy by using numerical simulations of synthetic turbulence made of a coherent field changing orientation on scale $l_c$, responsible for the wandering of magnetic field lines, and a power-law Kolmogorov-like turbulent spectrum on smaller scales, mimicking the pre-existing turbulence that provides resonant scattering, with independent normalizations. Since the main point here is to explore CR confinement when little or no parallel scattering is present, we only considered the case of a Kolmogorov spectrum without large scale magnetic field. When the two components have comparable strength the propagation of CRs is diffusive with a diffusion coefficient in agreement with previous work \citep{subedi2017charged,dundovic2020novel}. Interestingly, when the power in the large-scale component is dominant, low-energy particles diffuse with a pathlength that is roughly energy independent and of the order of $l_c$. When the small-scale turbulence is increased, the level of resonant scattering increases as well. This result implies that there should be at least one change in the energy dependence of the diffusion coefficient where the pathlength due to self-generated waves becomes smaller than $l_c$. In fact, as we discussed throughout the article, some level of resonant scattering due to a subdominant turbulent component of the field would also result is a similar effect. 

For values of the coherence length in the Galactic halo routinely adopted in the literature $l_c\sim 30$ pc \citep{Beck2015}, the diffusion pathlength of CRs as measured from secondary/primary ratios at $\lesssim$TeV energies becomes $\sim l_c$ in the $\sim$TeV energy range. Clearly, any power at small scales that may cause parallel scattering adds to this phenomenon, making the confinement time at such energies longer. We do not claim that the idealized situation investigated here describes CR Galactic confinement, but rather point out that there is a minimum level of CR diffusion even when there is no efficient parallel scattering due to resonant diffusion.

There is clearly a lot of work to be done to clarify the main processes responsible for CR confinement at energies $\gtrsim$ TeV, from exploring self-generation in non trivial models \cite[]{Dogiel2022} to characterizing the combined effect of FLRW (possibly with a range of values of $l_c$) with scattering due to different types of turbulent cascades, and finally investigating the effect of a large-scale homogeneous magnetic field. Nevertheless, it is important to realize that simulations can hardly be the final answer to all these questions, due to the limited range of scales that by construction these simulations can describe. More work on the basic principles of particle motion may serve as precious support to the limitations of numerical methods and viceversa simulations of MHD turbulence may provide invaluable tools to understand the essential ingredients to be accounted for in the theory of particle confinement \citep{gan2022existence}.

\section*{Acknowledgements}
The authors are grateful to the referee for several comments that helped us in improving the manuscript. Simulations presented here have been performed on the Newton cluster at the University of Calabria (Italy) and the work is supported by “Progetto STAR 2-PIR01 00008” (Italian Ministry of University and Research). The authors thank A. Marcowith for reading a preliminary version of this manuscript, and C. Evoli, O. Fornieri, W.H. Matthaeus, and S.S. Cerri for fruitful discussion. We also acknowledge the informal collaboration “Microphysics of Cosmic-Ray Observables” (MiCRO), led by S.S. Cerri, that facilitated discussions on the topic during the realisation of this work. 

\section*{Data Availability}
The data underlying this article will be shared on reasonable request to the corresponding author.



\bibliographystyle{mnras}
\bibliography{CRTransport_FLRWvsResonant} 






\bsp	
\label{lastpage}
\end{document}